\documentclass[prl,preprint,amsmath,amssymb]{revtex4}

\usepackage{graphicx}
\usepackage{dcolumn}
\usepackage{bm}

\begin{document}

\input epsf.sty



\title{Asymptotic and intermediate long-time behavior of nuclear free induction decays in polycrystalline solids and powders.}

\author{B. V. Fine}
\email{B.Fine@thphys.uni-heidelberg.de}
\affiliation{Institute for Theoretical
Physics, University of Heidelberg, Philosophenweg 19, 69120
Heidelberg, Germany}
\author{T. A. Elsayed}
\affiliation{Institute for Theoretical
Physics, University of Heidelberg, Philosophenweg 19, 69120
Heidelberg, Germany}
\author{E.  G. Sorte}
\affiliation{Department of
Physics, University of Utah, 115 South 1400 East, Salt Lake City,
Utah 84112-0830, USA}
\author{B.\ Saam}
\affiliation{Department of Physics,
University of Utah, 115 South 1400 East, Salt Lake City, Utah
84112-0830, USA}

\date{January 4, 2012}

\begin{abstract}
Free induction decay (FID) measured by nuclear magnetic resonance (NMR) in a polycrystalline solid is the isotropic average of the FIDs for individual single crystallites. It has been recently proposed theoretically and verified experimentally that the long-time behavior of single-crystal FIDs has the universal form of exponentially decaying sinusoidal oscillations. Polycrystalline averaging complicates the situation theoretically, while the available experimental evidence is also ambiguous. Exponentially decaying sinusoidal oscillations have been observed for $^{129}$Xe in polycrystalline solid xenon but not for $^{19}$F in the powder of CaF$_2$. In this paper, we present the first principles FID calculations for the powders of both CaF$_2$ and solid xenon. In both cases, the asymptotic long-time behavior has the expected form of exponentially decaying sinusoidal oscillations, which is determined by the single crystallite FID with the slowest exponential decay. However, this behavior  appears only at rather small values of the signal that have not yet been measured in experiments. At intermediate times accessible experimentally, a polycrystalline FID depends on the distribution of the exponential decay constants and oscillation frequencies for single crystallite FIDs. In  CaF$_2$, these parameters  are relatively broadly distributed, and as a result, the sinusoidal long-time oscillations become  somewhat washed out. In contrast,  the single crystallite parameters are more clustered in solid xenon, and, as a result, the experimentally observable range is characterized by well-defined oscillation frequency and exponential decay constant even though both of these parameters do not represent the true long-time behavior.  The above difference of the intermediate FID behavior originates from the difference of the crystal structures of solid xenon and CaF$_2$.

\end{abstract}
\pacs{76.60.Lz, 76.60.Es, 05.45.Gg, 03.65.Yz}


\maketitle

\section{Introduction}
\label{intro}

First principles calculations of the free induction decay (FID) measured by nuclear magnetic resonance (NMR) in solids is a long-standing theoretical problem\cite{Abragam-61} still lacking a controllable solution\cite{Abragam-61}. The most challenging aspect of this problem is the prediction of the long-time behavior of the FIDs. Recently some progress in this direction was made on the basis of the notion of microscopic chaos\cite{Fine-00,Fine-04,Fine-05}. Namely, it was predicted that the generic long-time behavior of FIDs in single crystals has the character of exponential decay with or without sinusoidal oscillations. In the most common case of magnetic dipolar interaction between nuclear spins,  the oscillatory regime is realized, and hence, the long-time FID behavior can be parameterized as
\begin{equation}
F(t) = A e^{- \gamma t} \hbox{cos}(\omega t + \phi),
\label{ltform}
\end{equation}
where $A$, $\gamma$, $\omega$ and $\phi$ are some constants whose values were not predicted.   It was only estimated\cite{Fine-04} that, generically, the values of  $\gamma$ and $\omega$  fall on the timescale of the spin-spin interactions often referred to as $T_2$.  It was also estimated that the long-time behavior (\ref{ltform}) becomes dominant after a time on the order of several times $T_2$ from the beginning of the FID.   The above predictions agree with the experimental\cite{Engelsberg-74,Sorte-11,Meier-11} and numerical\cite{Fabricius-97,Fine-03} results for quantum and classical spin systems.

The situation becomes somewhat more involved theoretically for  polycrystalline samples or crystal powders.  Different orientations of single crystallites in polycrystals/powders with respect to an external magnetic field imply different microscopic Hamiltonians, and hence different values of $\gamma$ and $\omega$, which in turn leads to the additional averaging over the oscillation frequencies. At sufficiently long times, the crystallites exhibiting the smallest value of $\gamma$ should start dominating the overall response, and, therefore, the well-defined frequency of these crystallites should also control the overall decay. We call the latter regime the asymptotic long-time behavior. It is to be distinguished from the intermediate behavior, which we define as the regime, when the individual crystallites have reached their respective long-time regimes but the asymptotic polycrystalline long-time behavior is not yet reached.  The challenge here is to understand how long the above transition to the asymptotic behavior takes, and what the intermediate behavior looks like.
It is, in particular, possible that the intermediate behavior exhibits a tentative ``washing out'' of the FID beats.

On the experimental side, the available facts about the long-time FID behavior in polycrystals/powders do not reveal a consistent picture. On the one hand, no well-defined long-time beats of form \eqref{ltform} have been observed in the CaF$_2$ powder (within the range limited by the experimental signal-to-noise ratio)\cite{Barnaal-66,Sorte-11}. On the other hand, in hyperpolarized solid xenon, which is supposedly polycrystalline, the experiments reveal well-defined beats of form (\ref{ltform}) appearing rather quickly\cite{Morgan-08,Sorte-11}.  

In the latter case, the situation is complicated by the fact that hyperpolarized solid xenon is prepared in convection cells\cite{Su-04} by first optically polarizing xenon gas\cite{Walker-1997} and then rapidly cooling it into a liquid phase and subsequently quenching the liquid into the solid phase. As a result, an uncertainty remains about the proper thermalization of the resulting solid.   In addition, the formation of crystal structure in solid xenon is controlled by the relatively weak van der Waals interaction, which is known to allow significant residual atomic motion\cite{Yen-1963} that further complicates the theoretical analysis. A related unclear issue is the strength of the exchange coupling between xenon nuclei.  

In this paper, we assume that hyperpolarized solid xenon samples investigated in Refs.\cite{Morgan-08,Sorte-11} can be described as polycrystalline fcc lattices of immobile nuclear spins coupled by magnetic dipole interaction. We perform the first principles calculations of $^{129}$Xe FID on the basis of the approximation procedure introduced in Refs.\cite{Fine-97,Fine-00}. We also perform the first principles $^{19}$F FID calculation for the powder of CaF$_2$, where $^{19}$F nuclei form a simple cubic lattice.  Our goal is to verify whether the above calculations are sufficient to explain why the well-defined beats of form (\ref{ltform}) were seen in polycrystalline solid xenon\cite{Morgan-08,Sorte-11} but not in CaF$_2$ powder\cite{Gade-69,Sorte-11}.

\section{Theoretical approximation scheme}
\label{theory}

We will use the approximation scheme for FID calculations that was introduced in Ref.\cite{Fine-97} with small modifications added in Ref.\cite{Fine-00}. This scheme is quite similar to the one introduced earlier in Ref.\cite{Becker-76}. Alternative attempts to calculate powder FIDs were made in Refs.\cite{Gade-69,Canters-72}. 

The approximation technique of Ref.~\cite{Fine-97} results in a very accurate description  of the extended initial behavior of single crystal FIDs in CaF$_2$. It also leads to the long-time behavior of form \eqref{ltform}, but with constants noticeably different from those observed experimentally (see below). As explained in Ref.\cite{Fine-04}, an accurate prediction of the parameters in Eq.(\ref{ltform}) is not expected here due to the oversimplified  nature of the approximation. We are, however, mainly interested in the qualitative question of the difference the solid xenon and the CaF$_2$ powders posed at the end of the preceding section.   Answering this question presumably depends on the qualitative differences in the distributions of $\gamma$ and $\omega$  for different orientations of single crystallites in the external magnetic field. The approximations used should, therefore, be adequate for detecting such differences, if they exist. 

In CaF$_2$, $^{19}$F nuclei are characterized by spin 1/2,  gyromagnetic ratio $\gamma_{\hbox{\scriptsize g}} = 25166.2\  \hbox{s}^{-1} \hbox{G}^{-1}$ and abundance $\nu = 1.0$.  These nuclei form simple cubic lattice with period $d= 2.723 \ \hbox{\AA}$ (at 293~K). For solid xenon, we perform the calculation for the fcc lattice with the nearest neighbor distance $d = 4.4 \ \hbox{\AA}$ and  abundance $\nu = 0.86$ of $^{129}$Xe nuclei.  This abundance is representative of the sample most studied in Refs.\cite{Morgan-08,Sorte-11}.  Other nuclear isotopes present in this xenon sample are assumed to be non-magnetic. (Here, in particular, we neglect the contribution of the magnetic isotope $^{131}$Xe, which has spin 3/2 with a smaller gyromagnetic ratio. Its abundance is  2 per cent in the sample analyzed.) The $^{129}$Xe nuclei have spin 1/2 with  gyromagnetic ratio $\gamma_{\hbox{\scriptsize g}} = 7452.11 \  \hbox{s}^{-1} \hbox{G}^{-1}$. 

We obtain the powder FID as the average over large number of single crystallite FIDs. The orientation of each crystallite in the external magnetic field is selected randomly.

For each crystallite, we calculate the FID as the infinite temperature correlation function\cite{Abragam-61}
\begin{equation}
F(t) = {
\hbox{Tr} \left \{e^{{i \over \hbar} {\cal H} t} \sum_n I_n^x   e^{-{i \over \hbar} {\cal H} t} \sum_m I_m^x \right\}
\over
\hbox{Tr} \left \{ \sum_n {I_n^x}^2    \right\}
}
\label{Fdef}
\end{equation}
 for the microscopic Hamiltonian of the truncated magnetic dipole  interaction in the Larmor rotating reference frame:
\begin{equation}
{\cal H} = \sum_{m<n} J_{mn} \left[
I_m^z I_n^z - {1 \over 2} (I_m^x I_n^x + I_m^y I_n^y)
\right],
\label{H}
\end{equation}
where $m$ and $n$ are the lattice site indices, $I_m^{\delta}$ is the operator of the $\delta$th ($x$, $y$, or $z$) component of the $m$th nuclear spin 1/2 with the $z$-axis chosen along the direction of the external static magnetic field, and $J_{mn}$ are the coupling constants given by
\begin{equation}
J_{mn} = {\gamma^2_{\hbox{\scriptsize g}} \hbar^2 (1 - 3 \hbox{cos}^2 \theta_{mn}) \over  |{\mathbf r}_m - {\mathbf r}_n|^3}.
\label{J}
\end{equation}
Here, ${\mathbf r}_m$ is the position vector of the $m$th nucleus, and $\theta_{mn}$ is the angle between vector $({\mathbf r}_m - {\mathbf r}_n)$ and the $z$-axis.

Extending the approximation scheme of Refs.\cite{Fine-97,Fine-00} to the case of isotopic abundance $\nu <1$, we obtain the FID function $F(t)$ as the numerical solution of the following integral equation:
\begin{equation}
F(t) = g(t) + \alpha \int_0^t F(t-t^{\prime}) {d g(t^{\prime}) \over d t^{\prime}} d t^{\prime},
\label{Fint}
\end{equation}
where
\begin{equation}
g(t) = \hbox{exp} \left[-{1 \over 2} M_{2f} (\eta t)^2 \right]
\prod_n \left[
1 - \nu + \nu \hbox{cos}\left( {3 \over 4 \hbar} J_{mn} \eta t \right)
\right],
\label{g}
\end{equation}
\begin{equation}
\alpha = {
{M_{4g} \over M_{2g}^2} - {M_4 \over M_2^2}
\over
{M_4 \over M_2^2} - 1
},
\label{alpha}
\end{equation}
\begin{equation}
\eta = \sqrt{{M_2 \over (1+\alpha)(M_2 + M_{2f})}},
\label{eta}
\end{equation}
\begin{equation}
M_2 \equiv - \left. {d^2 F \over d t^2} \right|_{t=0} = {9 \nu \over 16 \hbar^2} \sum_n J_{mn}^2,
\label{M2}
\end{equation}
\begin{equation}
M_4 \equiv  \left. {d^4 F \over d t^4} \right|_{t=0} =
 {81 \over 256 \; \hbar^4} \left\{
{7 \over 3} \nu^2  \sum_{n,p}^{n \neq p} J_{mn}^2  J_{mp}^2 
+
{2 \over 3} \nu^2  \sum_{n,p}^{n \neq p} J_{mn}  J_{mp}  J_{np}^2
+
\nu \sum_n J_{mn}^4
\right\},
\label{M4}
\end{equation}
\begin{equation}
M_{2f} = {2 \over 7} M_2
\label{M2f}
\end{equation}
(see also \cite{Prefactor}),
\begin{equation}
M_{2g} \equiv - \left. {d^2 g \over d t^2} \right|_{t=0} = \eta^2 (M_{2f} + M_2),
\label{M2g}
\end{equation}
\begin{equation}
M_{4g} \equiv  \left. {d^4 g \over d t^4} \right|_{t=0} = \eta^4 \left[
3(M_2 + M_{2f})^2  +  {81 \over 256 \; \hbar^4}  (\nu - 3 \nu^2) \sum_n J_{mn}^4
\right].
\label{M4g}
\end{equation}
The initial condition for Eq.(\ref{Fint}) is $F(0) = 1$. 
Parameter $\alpha$ given by Eq.\eqref{alpha} does not depend on $\eta$. Therefore, one can first set $\eta = 1$, then calculate $\alpha$ and finally use Eq.\eqref{eta} to calculate the actual value of $\eta$.

\section{Results and discussion}
\label{results}

\begin{figure} \setlength{\unitlength}{0.1cm}

\begin{picture}(100, 204) 
{ 
\put(0, 136){ \epsfxsize= 3.3in \epsfbox{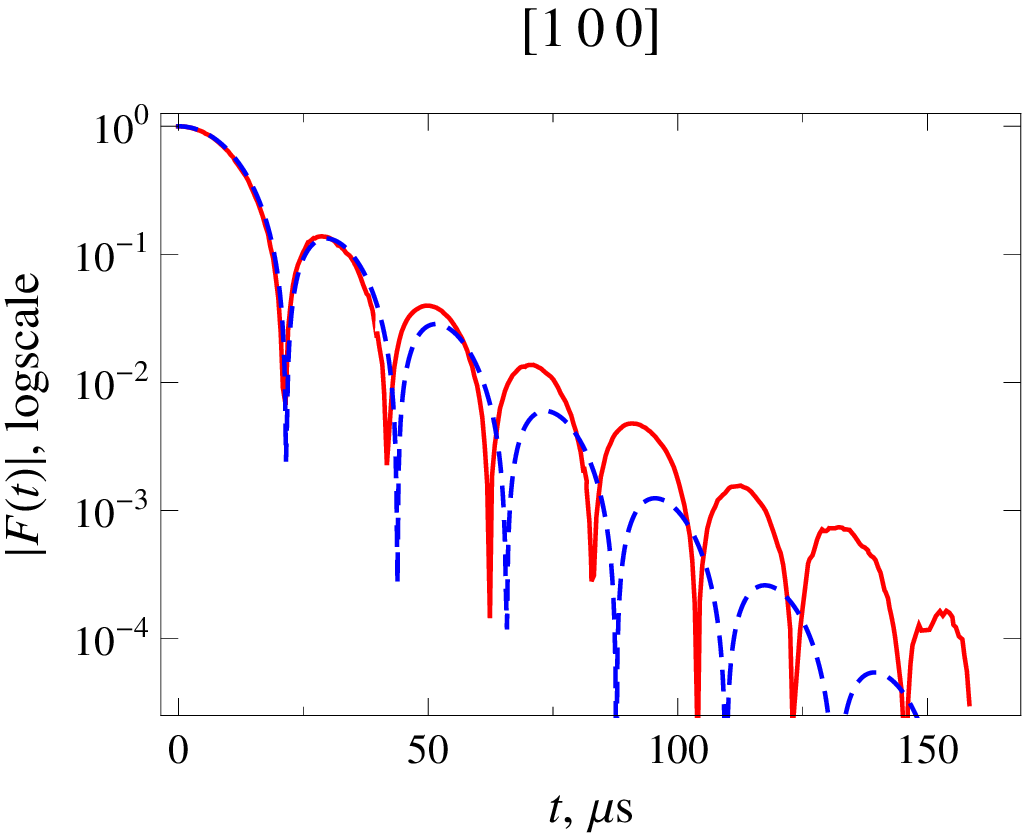} }
\put(0, 67){ \epsfxsize= 3.3in \epsfbox{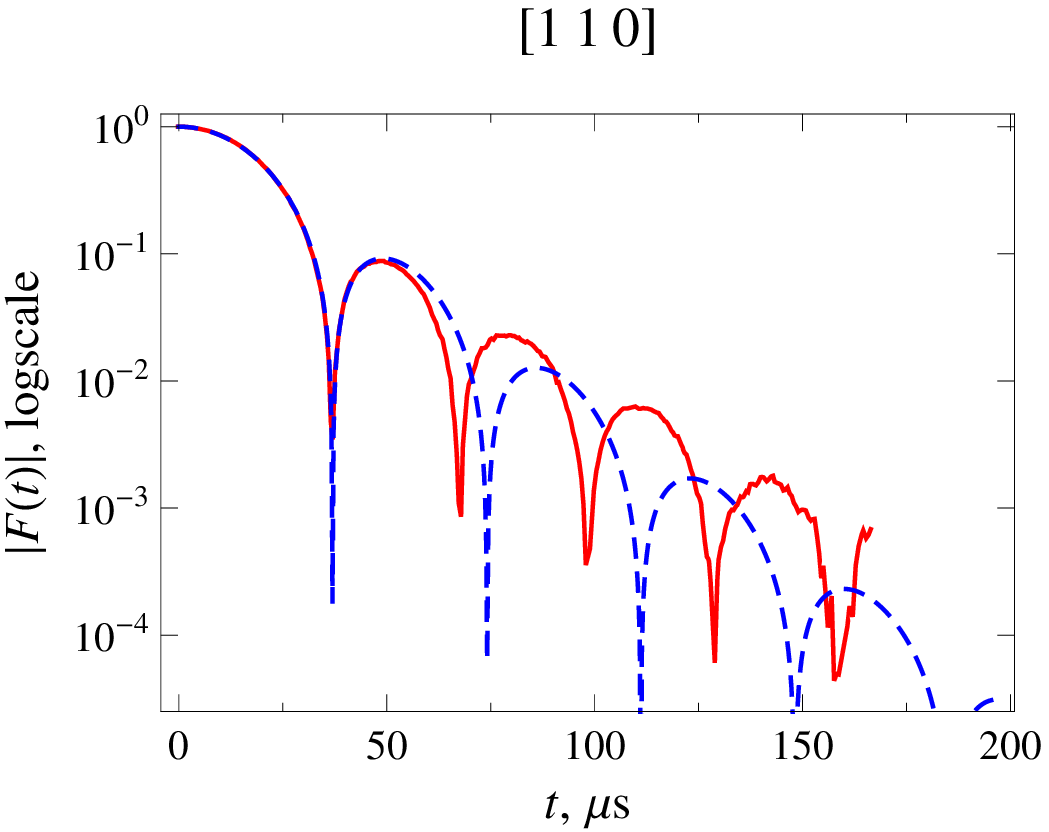} }
\put(0, -2){ \epsfxsize= 3.3in \epsfbox{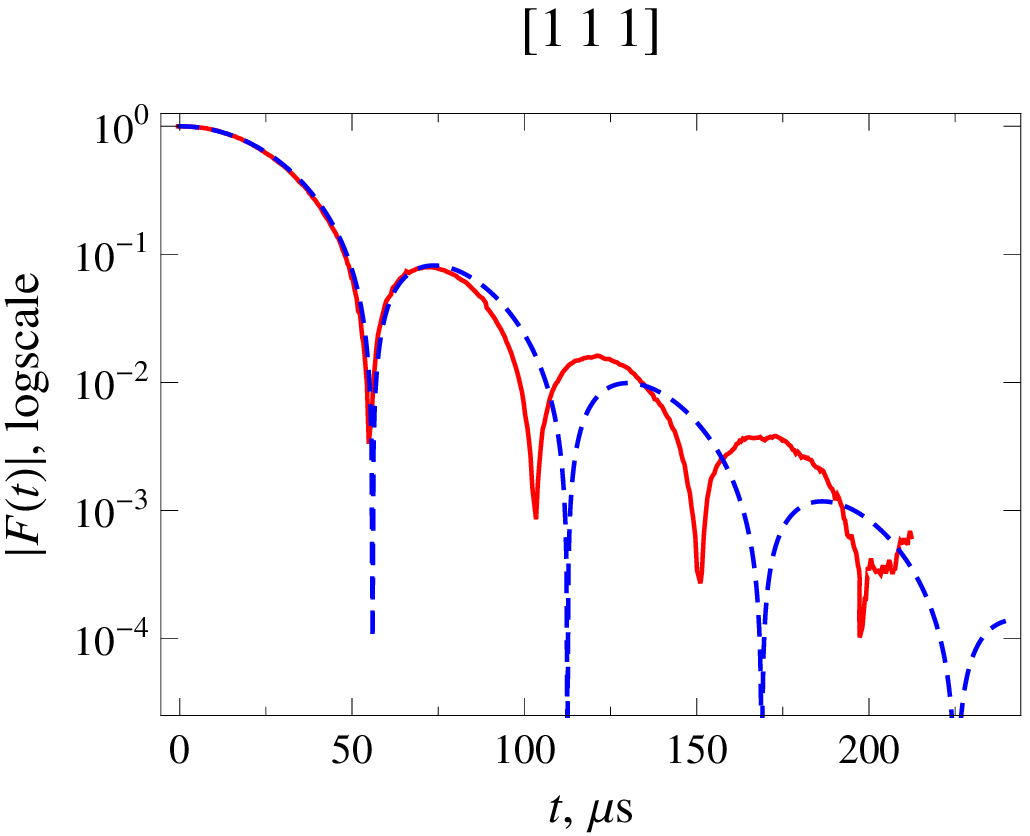} }
\put(49, 32){ \epsfxsize= 1.3in \epsfbox{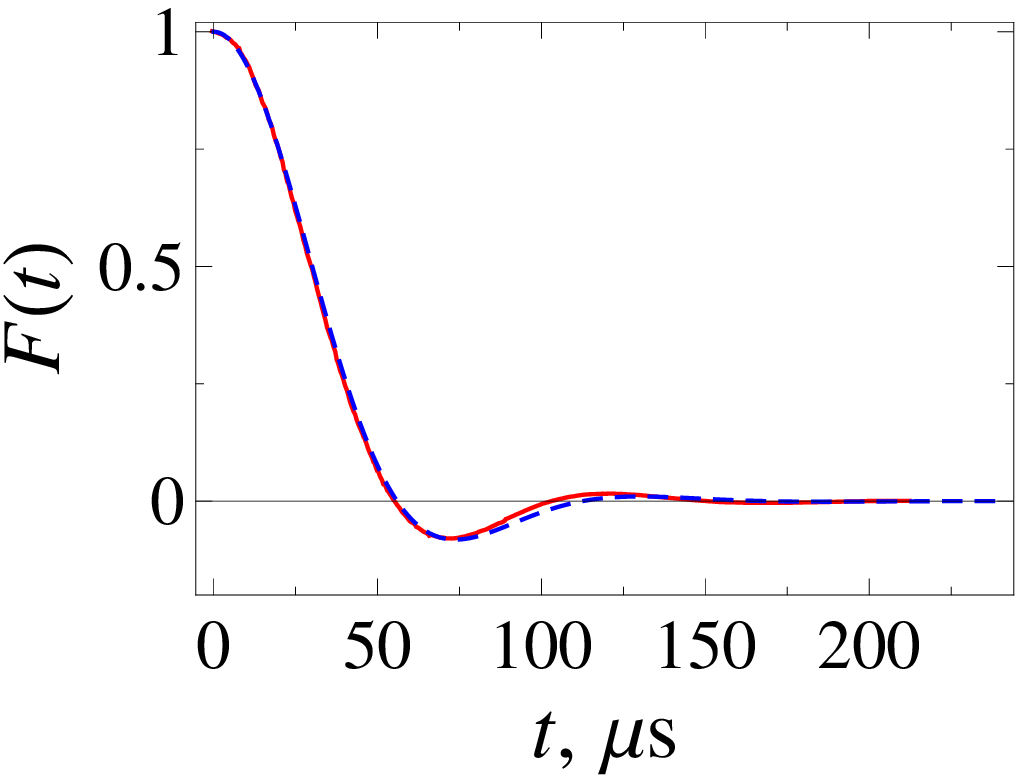} }
\put(49, 101){ \epsfxsize= 1.3in \epsfbox{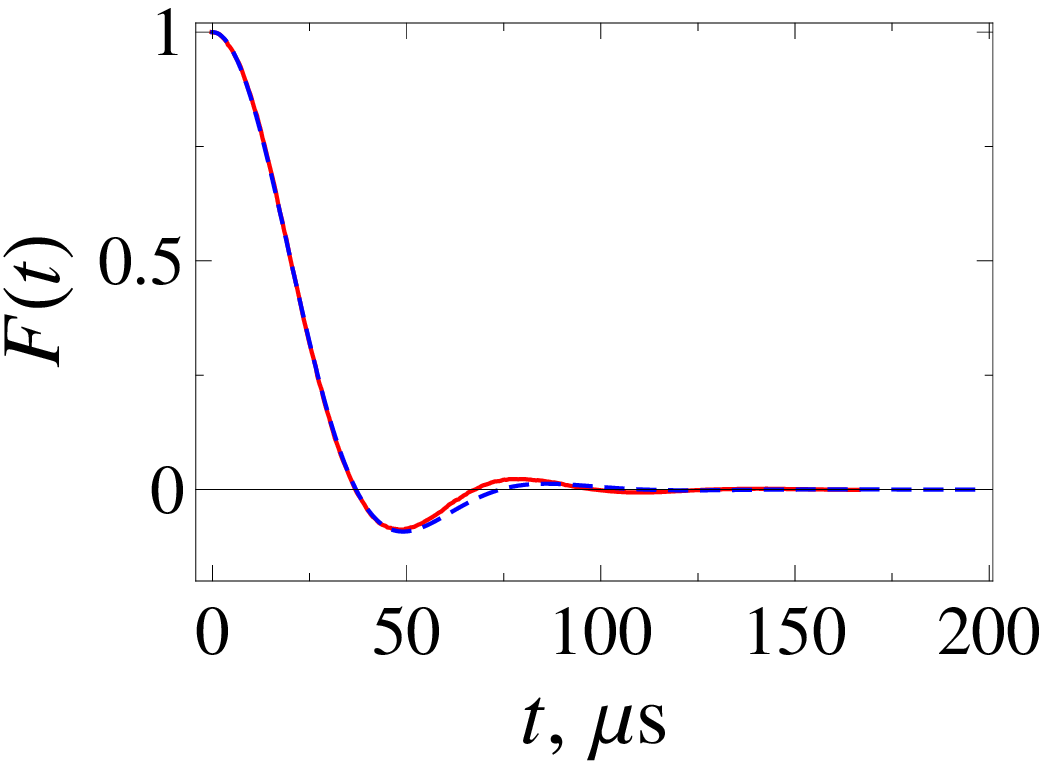} }
\put(49, 170){ \epsfxsize= 1.3in \epsfbox{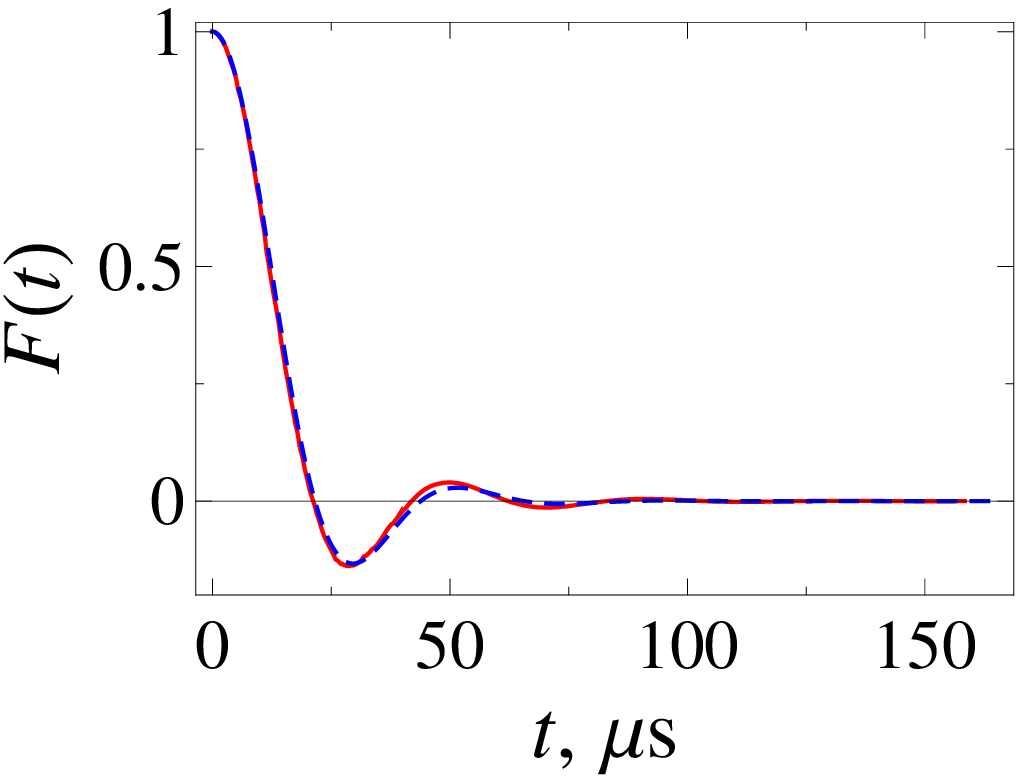} }
\put(0,61){ \textsf{\Large (c)} }
\put(0,128){ \textsf{\Large (b)} }
\put(0,195){ \textsf{\Large (a)} }
}
\end{picture} 
\caption{(Color online) Single crystal FIDs for CaF$_2$. The directions of the external magnetic fields are indicated above the plots. Solid red lines represent the experimental results of Engelsberg and Lowe\cite{Engelsberg-74}. Dashed blue lines represent the result of theoretical calculations based on Eq.(\ref{Fint}). Main panels contain semi-logarithmic plots. 
Insets --- linear plots.
} 
\label{fig-caf2-Cryst} 
\end{figure}

In order to illustrate the performance of the above approximation scheme, we show in Fig.~\ref{fig-caf2-Cryst} the results of the calculations  of the initial and the long-time behavior of single crystal FIDs in CaF$_2$ for three directions of the external magnetic field. While the linear plots (insets) in each panel of Fig.~\ref{fig-caf2-Cryst} illustrate that the overall agreement of the theoretical and the experimental curves is very good, the semilog plots (main panels) amplify the discrepancy in the long-time tails. The comparison of the theoretical and the experimental values of $\gamma$ and $\omega$ for the long-time fits of form (\ref{ltform}) is presented in Table~\ref{tab-cryst}. It indicates a typical discrepancy of about 20 percent.


\begin{table}[t]
\begin{center}
\begin{tabular}{|c||c|c||c|c||c|c|} \hline
  & \multicolumn{2}{c||}{[100]}  & \multicolumn{2}{c||}{[110]}   & \multicolumn{2}{c|}{[111]}  \\
\cline{2-7}
 & $\gamma$, $\mu$s$^{-1}$  &  $\omega$, $\mu$s$^{-1}$ &  $\gamma$, $\mu$s$^{-1}$  &  $\omega$, $\mu$s$^{-1}$ & $\gamma$, $\mu$s$^{-1}$  &  $\omega$, $\mu$s$^{-1}$   \\
\hline
exp. & 0.054 & 0.156 & 0.042 & 0.103 & 0.029 & 0.066 \\
\hline
th. & 0.071 & 0.143 & 0.054 & 0.085 & 0.038 & 0.056 \\
 \hline
\end{tabular}
\end{center}
\caption{Table summarizing the experimental and theoretical values of parameters $\gamma$ and $\omega$ for the single crystal CaF$_2$. The parameters are obtained by fitting the long-time tails of the FIDs presented in Fig.~\ref{fig-caf2-Cryst} by Eq.(\ref{ltform}). 
}
\label{tab-cryst}
\end{table}

\begin{figure} \setlength{\unitlength}{0.1cm}
\begin{picture}(100, 185) 
{ 
\put(0, 132){ \epsfxsize= 3.3in \epsfbox{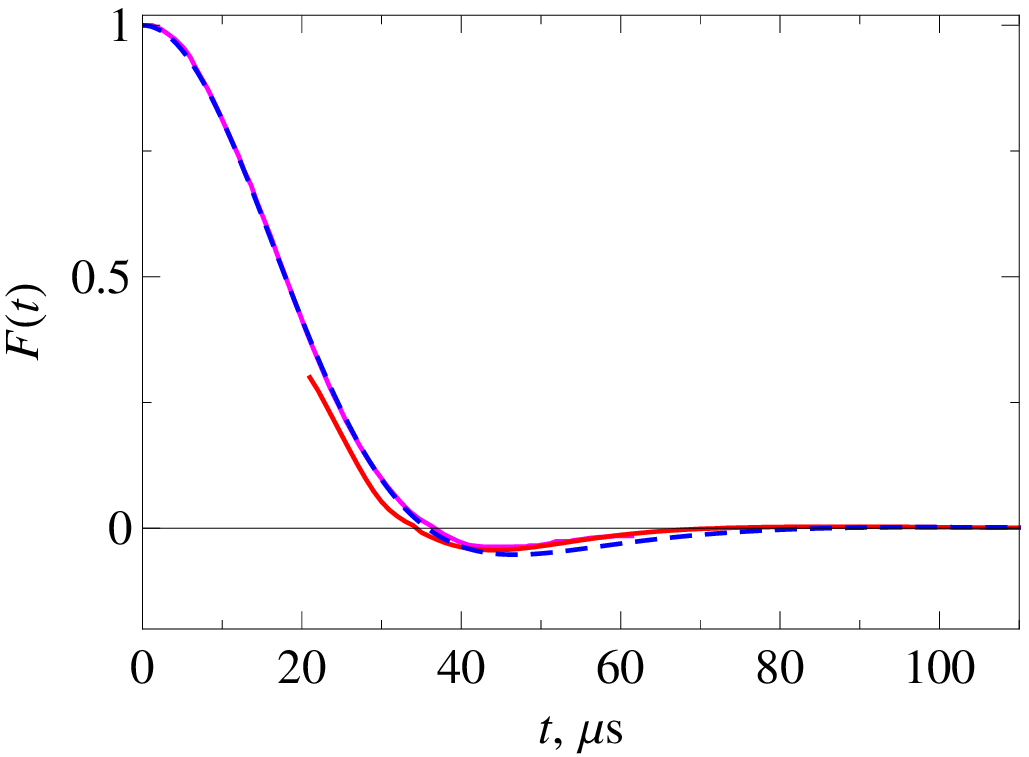} }
\put(0, 65){ \epsfxsize= 3.3in \epsfbox{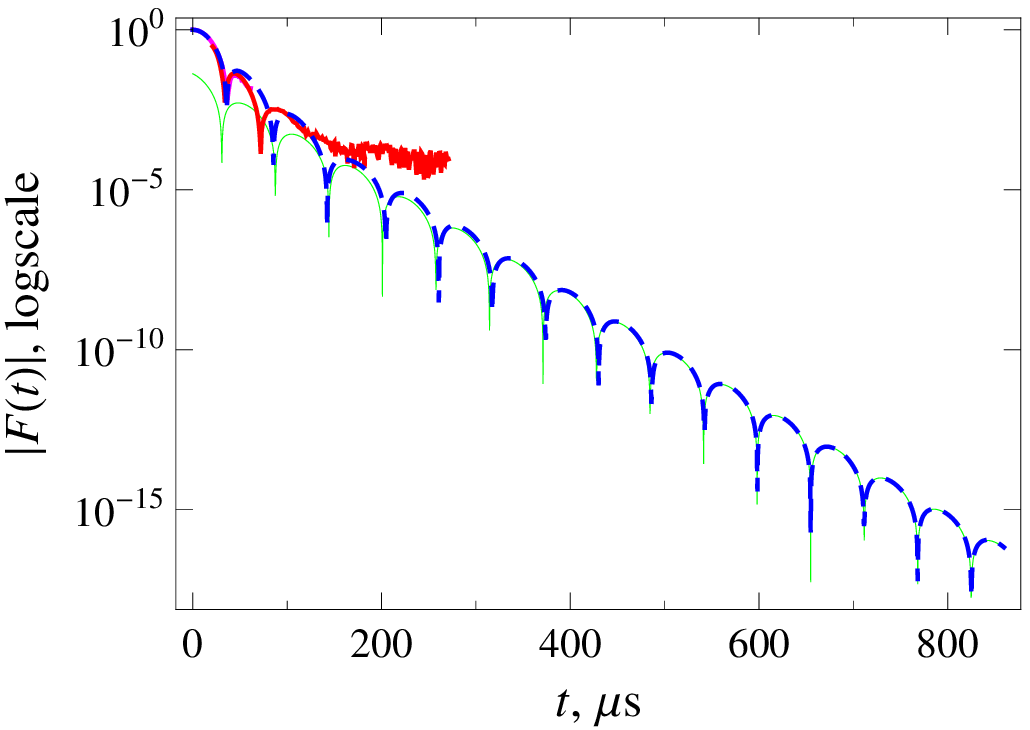} }
\put(0, -2){ \epsfxsize= 3.3in \epsfbox{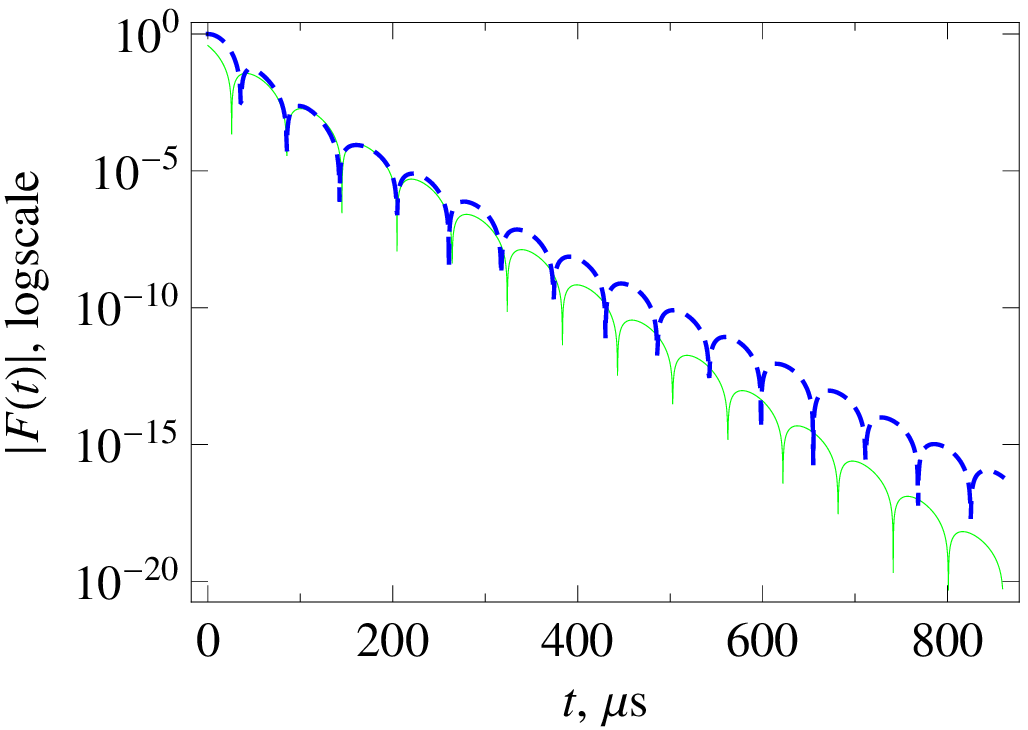} }
\put(0,61){ \textsf{\Large (c)} }
\put(0,128){ \textsf{\Large (b)} }
\put(0,195){ \textsf{\Large (a)} }
}
\end{picture} 
\caption{(Color online) $^{19}$F  FID for CaF$_2$ powder. The solid red line represents the experiment of Ref.\cite{Sorte-11}. The solid magenta line represents the earlier experiment of Barnaal and Lowe\cite{Barnaal-66}. (The data points are actually extracted from Ref.\cite{Gade-69}.) The dashed blue line represents the theoretical calculation described in the text. (a) Linear plot. (b) Semi-logarithmic plot with the thin green line representing the fit of form\eqref{ltform} for the asymptotic long-time behavior of the theoretical FID. (c) Semi-logarithmic plot with an attempt to fit the intermediate FID behavior by formula \eqref{ltform}.
} 
\label{fig-caf2-powder} 
\end{figure}

\begin{figure} \setlength{\unitlength}{0.1cm}
\begin{picture}(100, 185) 
{ 
\put(0, 132){ \epsfxsize= 3.3in \epsfbox{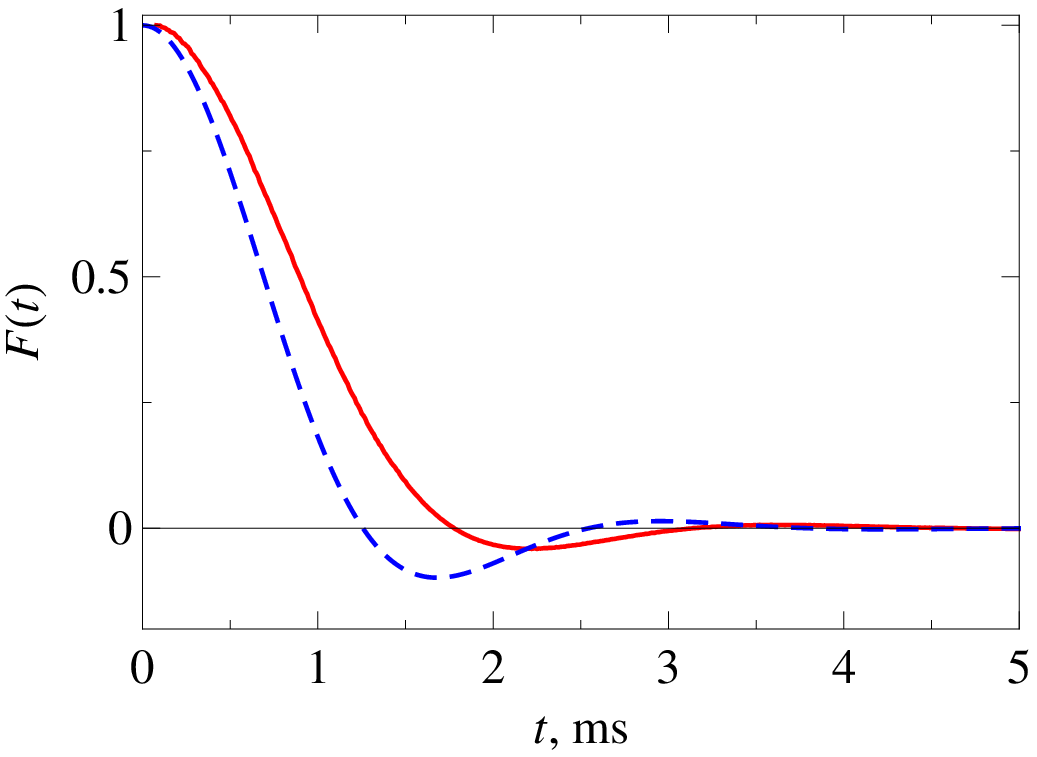} }
\put(0, 65){ \epsfxsize= 3.3in \epsfbox{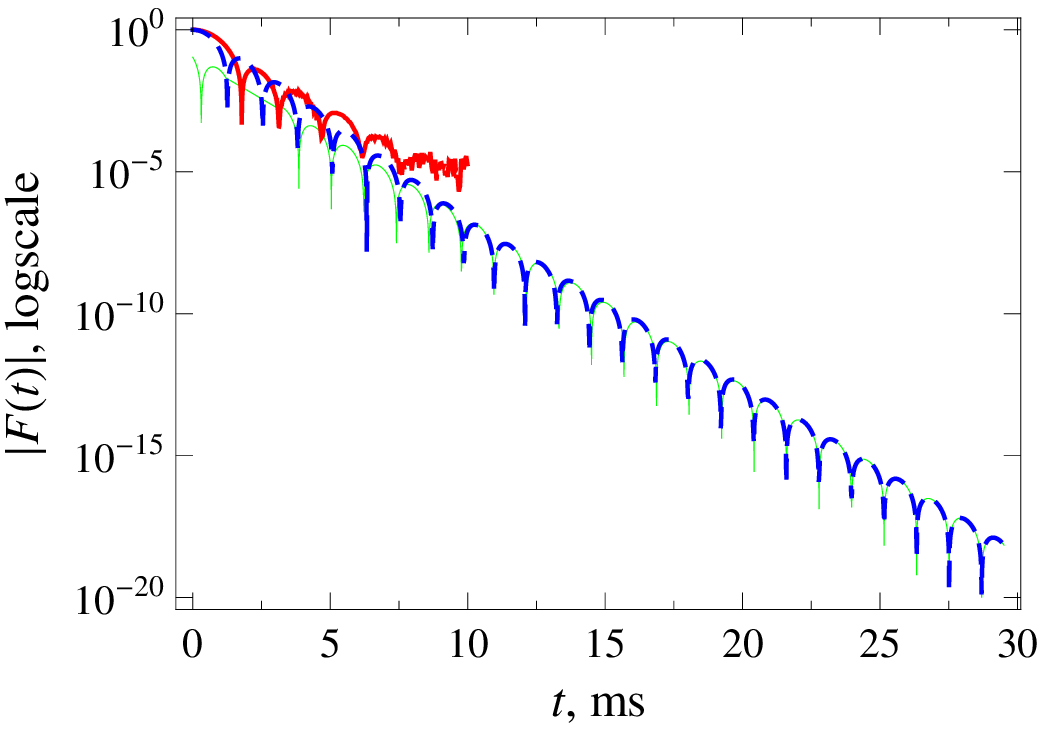} }
\put(0, -2){ \epsfxsize= 3.3in \epsfbox{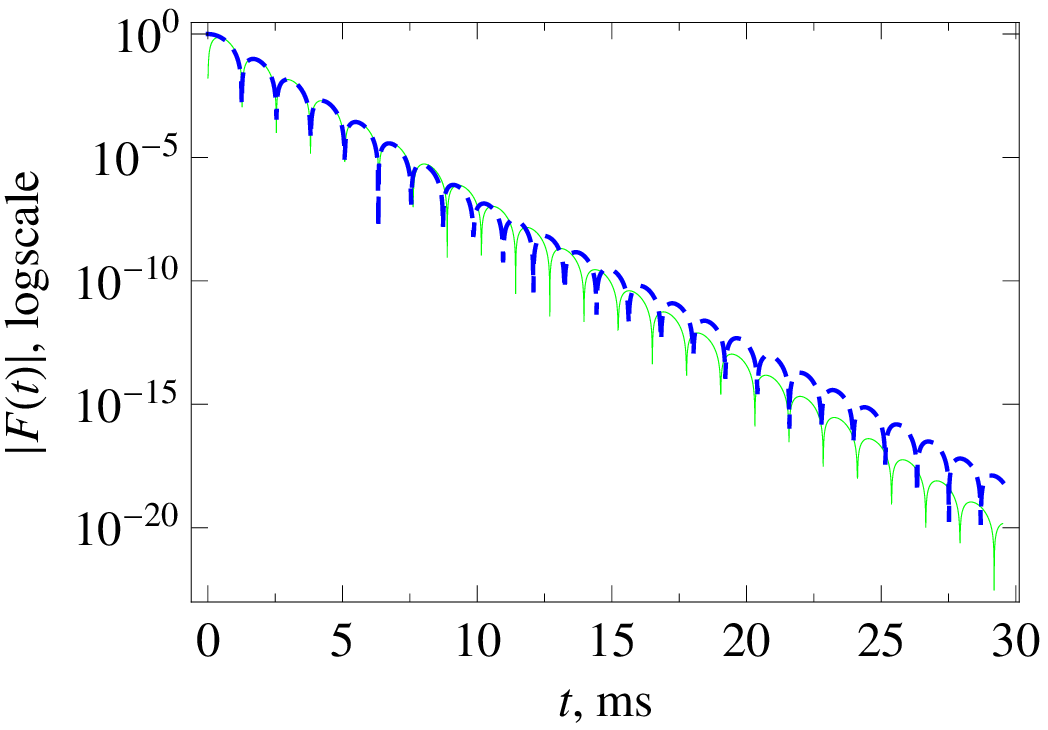} }
\put(0,61){ \textsf{\Large (c)} }
\put(0,128){ \textsf{\Large (b)} }
\put(0,195){ \textsf{\Large (a)} }
}
\end{picture} 
\caption{(Color online) $^{129}$Xe FID for polycrystalline xenon. The solid red line represents the experiment of Ref.\cite{Sorte-11}, the dashed blue line is the theoretical calculation described in the text. (a) Linear plot. (b) Semi-logarithmic plot with the thin green line representing the fit of form\eqref{ltform} to the asymptotic long-time behavior of the theoretical FID. (c) Semi-logarithmic plot with an attempt to fit the intermediate FID behavior by formula \eqref{ltform}.
} 
\label{fig-xenon} 
\end{figure}

The calculated FIDs for the CaF$_2$ powder and for polycrystalline solid xenon are shown in Figs.~\ref{fig-caf2-powder} and \ref{fig-xenon}. These FIDs were obtained as the average over 1000 randomly oriented single crystallites. 


\begin{table}[t]
\begin{center}
\begin{tabular}{|c||c|c||c|c|} \hline
  & \multicolumn{2}{c||}{CaF$_2$}  & \multicolumn{2}{c|}{solid Xe}     \\
\cline{2-5}
 & $\gamma$, $\mu$s$^{-1}$  &  $\omega$, $\mu$s$^{-1}$ &  $\gamma, \;$ms$^{-1}$  &  $\omega, \;$ms$^{-1}$    \\
\hline
theory (asymptotic) & 0.040  &  0.055 & 1.35  &  2.66  \\
\hline
theory (intermediate) & 0.050  & 0.053   & 1.55  &  2.48   \\
 \hline
experiment & -  & -  & 1.25  & 2.06   \\
\hline
\end{tabular}
\end{center}
\caption{Table summarizing the long-time parameters  $\gamma$ and $\omega$ corresponding to the theoretical calculations and experiments for CaF$_2$ powder and polycrystalline solid xenon.  The numbers in the ``theory~(asymptotic)'' row are obtained from fitting Eq.(\ref{ltform}) to the true theoretical long-time behavior as exhibited in Figs.~\ref{fig-caf2-powder}(b) and \ref{fig-xenon}(b).  The numbers in the ``theory~(intermediate)'' row are obtained from fitting the intermediate behavior of the theoretical FIDs in Figs.~\ref{fig-caf2-powder}(c) and \ref{fig-xenon}(c) by Eq.(\ref{ltform}).  The experimental numbers for solid xenon are cited from Ref.\cite{Sorte-11}. 
}
\label{tab-powders}
\end{table}

Let us first examine the discrepancies between the theoretical calculations and the experimental curves. For the CaF$_2$ powder, the discrepancies appear starting from the intermediate section of the FID. We believe that these discrepancies are due to the limitations of the theoretical approximation scheme based on Eq.(\ref{Fint}). On the other hand, the discrepancy for the solid xenon appears from the very beginning of the FID. It is related to the fact that the theoretical and the experimental values of the second moment $M_2$ are different from each other ($2.64 \ \hbox{ms}^{-2}$ and $1.6 \ \hbox{ms}^{-2}$, respectively). Since the theoretical value of $M_2$ is the input rather than the output parameter for the theoretical approximation scheme, the above discrepancy indicates the inadequacy of our initial assumptions about either the form or the parameters of the Hamiltonian (\ref{H}). It may be related to the insufficient thermalisation  and/or atomic motions in the quenched solid xenon samples\cite{Sorte-11}. Leaving this discrepancy to be investigated in a later experimental work, below we focus on the outcome of the theoretical calculation and examine the differences between the long-time FID behavior for CaF$_2$ powder and polycrystalline solid xenon.

Figures~\ref{fig-caf2-powder}(b) and \ref{fig-xenon}(b)  include fits of the true theoretical long-time behavior to the asymptotic formula \eqref{ltform}, while Figs.~\ref{fig-caf2-powder}(c) and \ref{fig-xenon}(c) attempt to fit the middle section of the theoretical FIDs with Eq.(\ref{ltform}). In CaF$_2$,  the intermediate FID behavior is not well described by Eq.\eqref{ltform}. At the same time, the asymptotic long-time behavior becomes pronounced relatively quickly --- after about three beats.
On the contrary, the intermediate behavior of the solid xenon FID is well described by Eq.\eqref{ltform}, which covers about 6 beats and 5 orders of magnitude, while the asymptotic long-time behavior emerges only at relatively late times and small values of FID. 

\begin{figure} \setlength{\unitlength}{0.1cm}
\begin{picture}(100, 180) 
{ 
\put(0, 93){ \epsfxsize= 3.3in \epsfbox{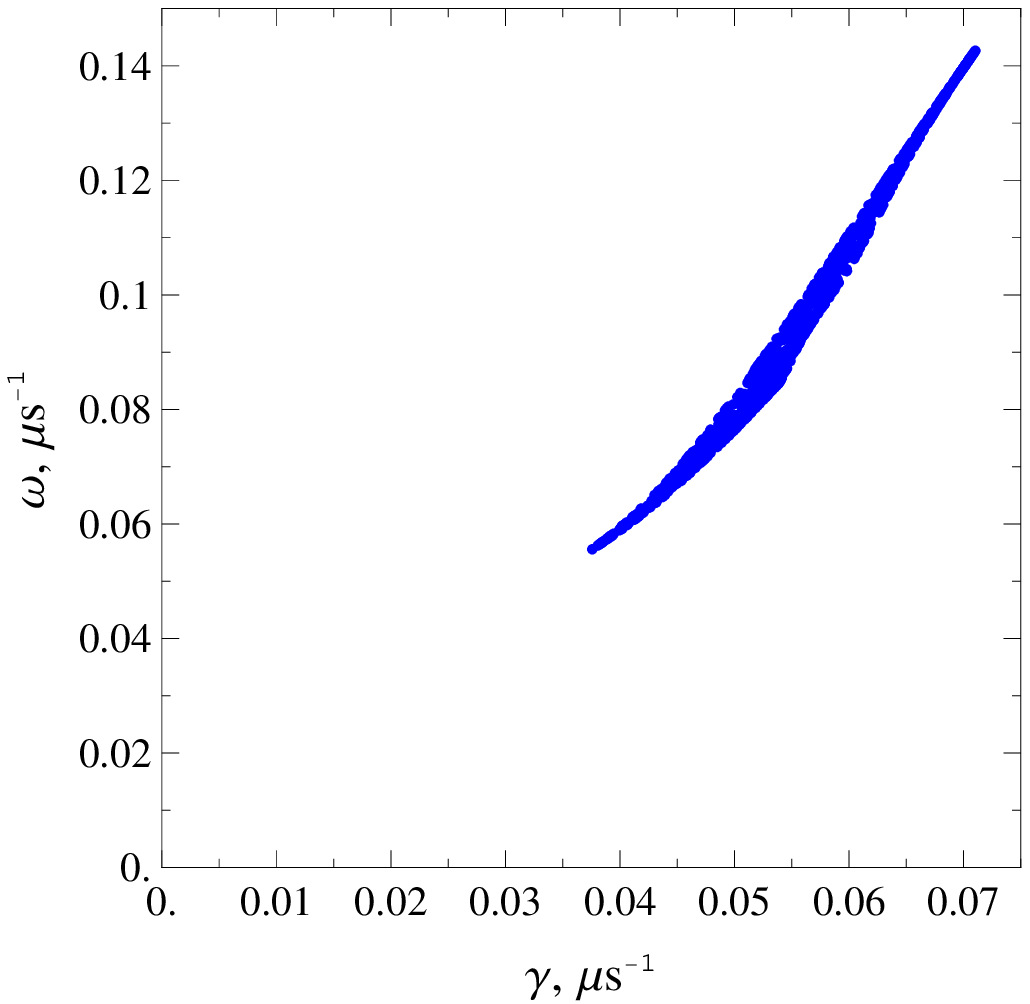} }
\put(10, 3){ \epsfxsize= 2.9in \epsfbox{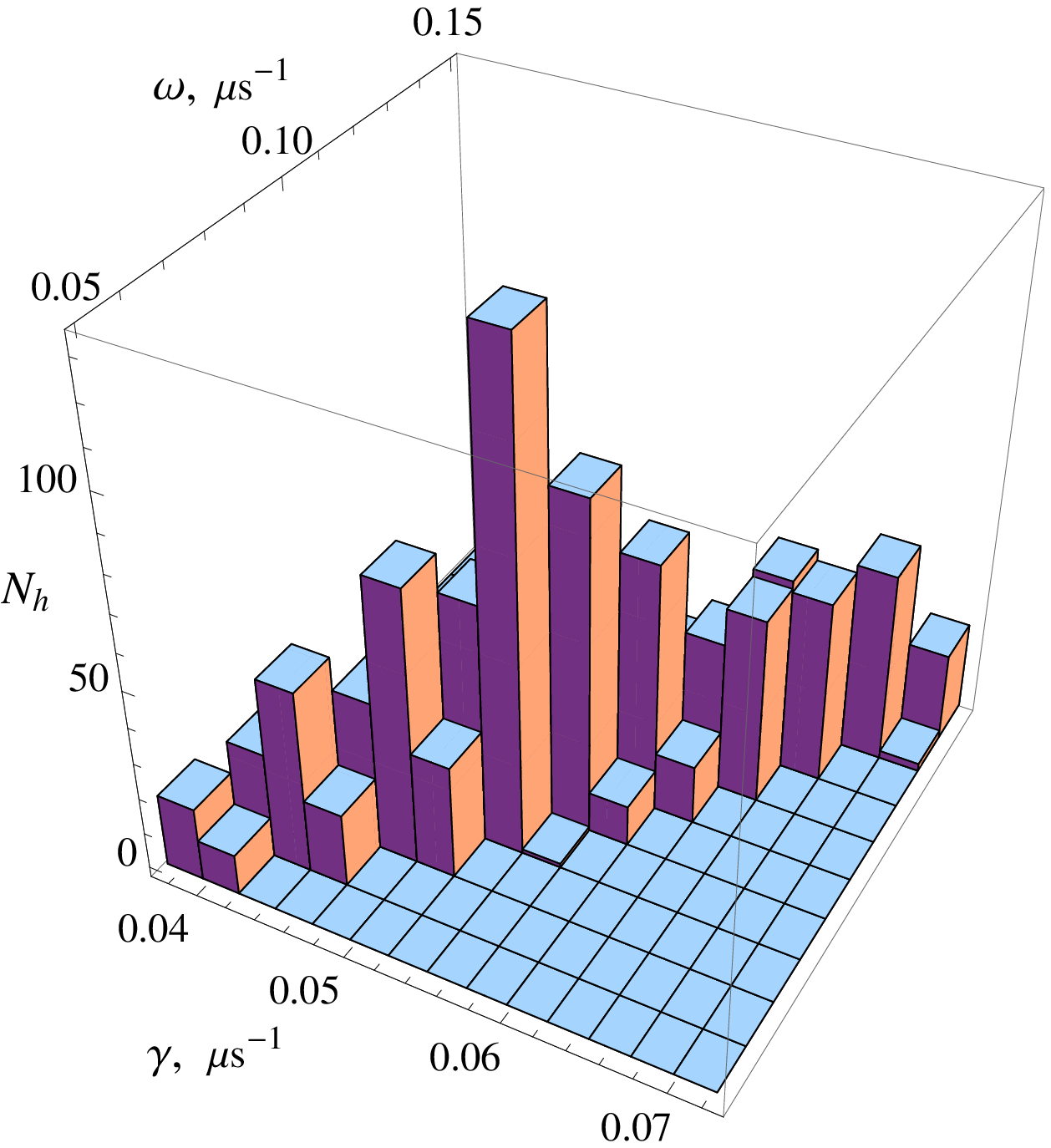} }
\put(0,175){ \textsf{\Large (a)} }
\put(0,80){ \textsf{\Large (b)} }
}
\end{picture} 
\caption{(Color online) (a) Theoretical values of parameters $\gamma$ and $\omega$ for 1000 randomly chosen single crystal orientations of CaF$_2$. The blue shape appearing in the plot consists of 1000 points. Each point represents a pair of values $(\gamma, \omega)$ for one single crystal orientation. (b) Histogram $N_h(\gamma, \omega)$ of all sampled points in (a).  
} 
\label{fig-gmom-caf2} 
\end{figure}

\begin{figure} \setlength{\unitlength}{0.1cm}
\begin{picture}(100, 180) 
{ 
\put(0, 93){ \epsfxsize= 3.3in \epsfbox{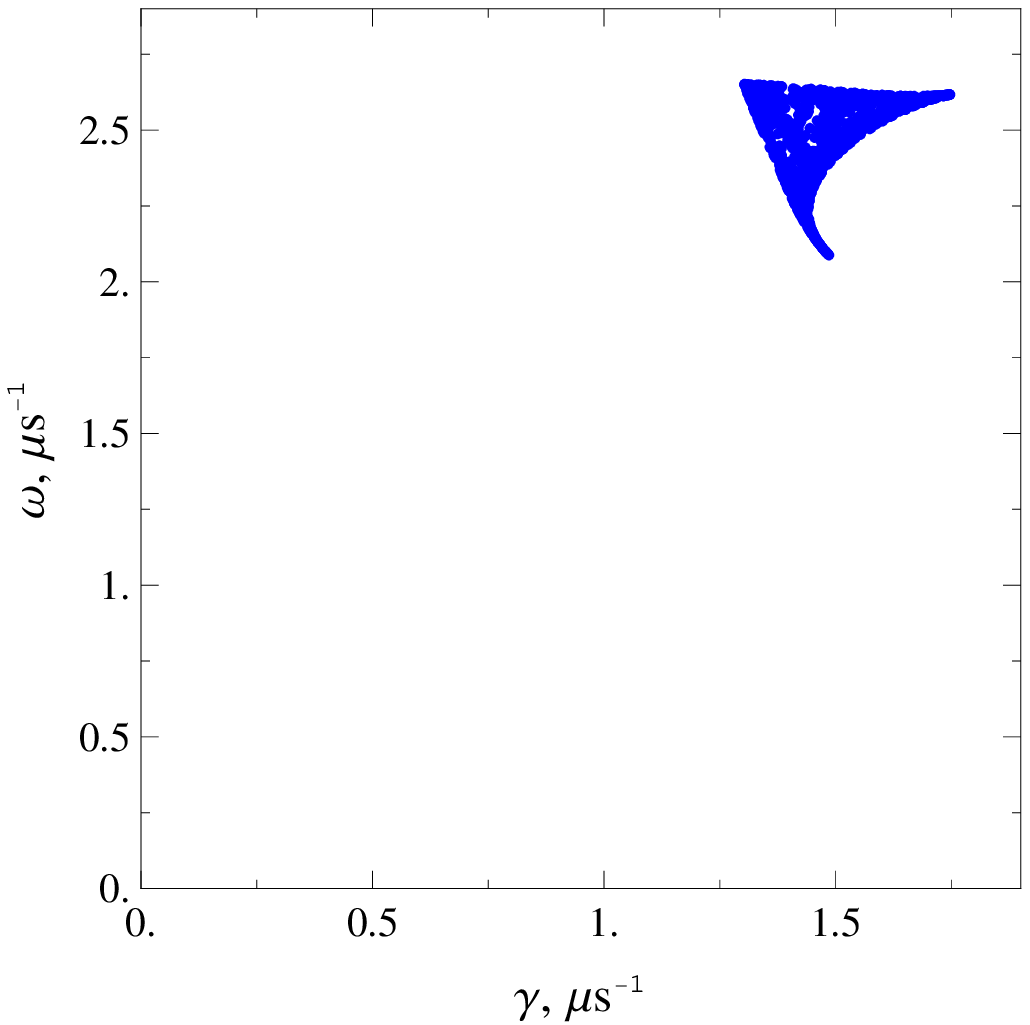} }
\put(10, 3){ \epsfxsize= 2.9in \epsfbox{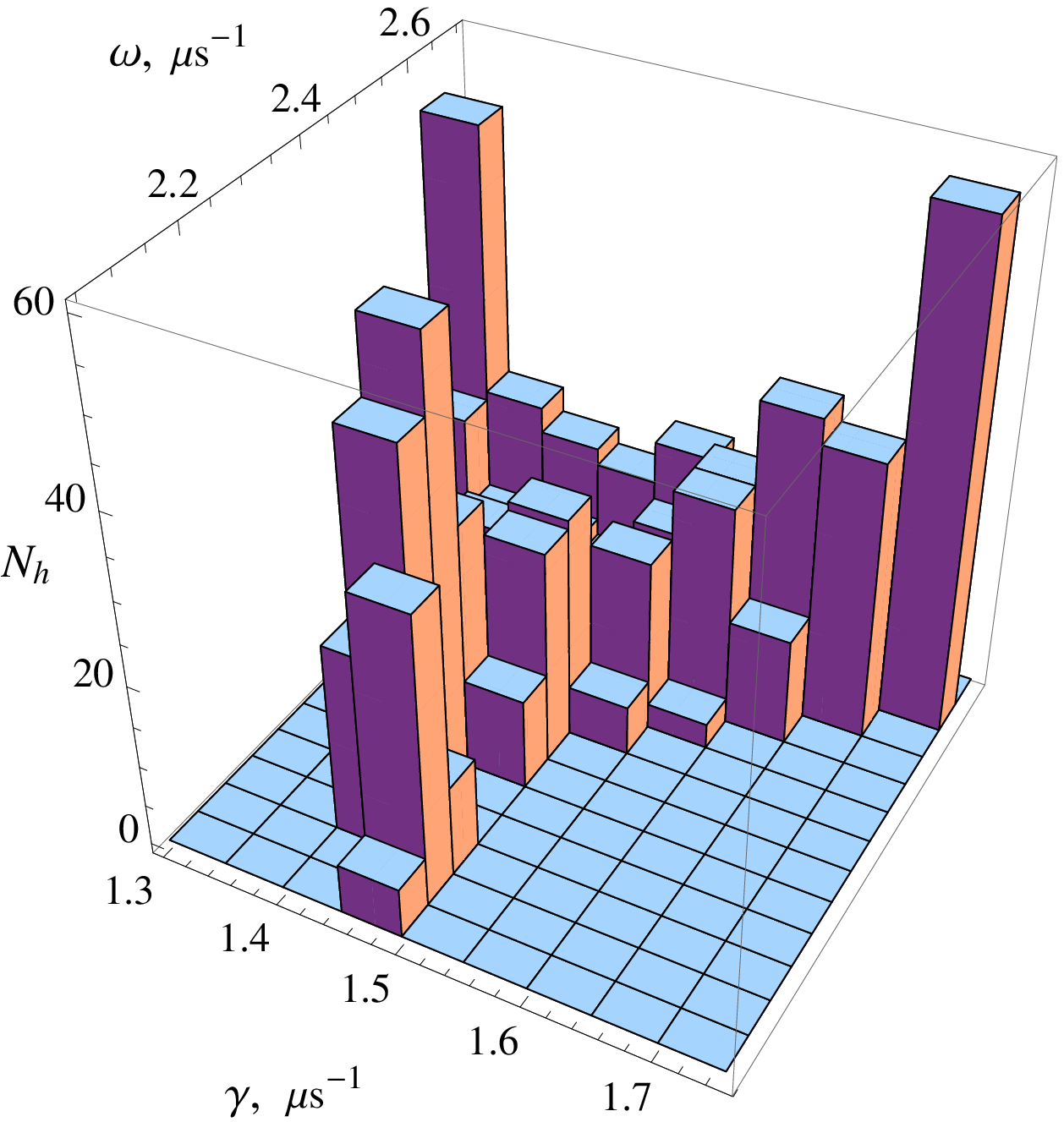} }
\put(0,175){ \textsf{\Large (a)} }
\put(0,80){ \textsf{\Large (b)} }
}
\end{picture} 
\caption{(Color online) (a) Theoretical values of parameters $\gamma$ and $\omega$ for 1000 randomly chosen single crystal orientations of solid xenon. The blue shape appearing in the plot consists of 1000 points. Each point represents a pair of values $(\gamma, \omega)$ for one single crystal orientation. (b) Histogram $N_h(\gamma, \omega)$ of all sampled points in (a).  
} 
\label{fig-gmom-xenon} 
\end{figure}

It is expected that the behavior of the middle section of the FIDs is controlled by the typical single crystallite values of  $\gamma$,  while the true long-time behavior is controlled by the crystallites with the smallest value of $\gamma$.
In order to clarify this issue further, we present in Figs.~\ref{fig-gmom-caf2}(a) and \ref{fig-gmom-xenon}(a) the theoretical values of the long-time parameters $\gamma$ and $\omega$ for the single crystallites included in the powder average, while Figs.~\ref{fig-gmom-caf2}(b) and \ref{fig-gmom-xenon}(b) show the histogram of the resulting points.


One can now appreciate the qualitative difference between the powder  of simple cubic crystallites and the powder of fcc crystallites. The long-time parameters $\gamma$ and $\omega$ for the simple cubic lattice are much broader and increase or decrease roughly proportionally to each other.  Therefore, the typical values of $\gamma$ and $\omega$   are sufficiently different from those representing the asymptotic decay. The poor performance of the middle section fit is in large part due to the larger difference of frequency $\omega$ between a typical value and the asymptotic long-time value. 

On the contrary, all possible values of $\gamma$ and $\omega$ are more clustered for the fcc polycrystal and do not exhibit much of a systematic dependence on each other. As a result, the typical value of $\gamma$ is rather close to the true long-time value. This explains why the fit~\eqref{ltform} to the intermediate FID behavior works so well over an extended time interval.

The more clustered behavior of parameters $\gamma$ and $\omega$ for the fcc powder was, in fact, expected. The differences in $\gamma$ and $\omega$ originate from the differences in the truncated Hamiltonians for different orientations of the magnetic field with respect to single crystallites. The orientation-dependent differences are expected to be smaller for the fcc lattice, because the fcc lattice is in a sense more isotropic:  each lattice site has 12 nearest neighbors as opposed to 6 nearest neighbors in the case of simple cubic lattice. The 12-neighbor environment is obviously more isotropic than 6-neighbor environment. The higher sensitivity of the simple cubic lattice to different orientations of the magnetic field can be illustrated by the example of the  magnetic field oriented along [111] crystal direction, in which case the coupling constants~\eqref{J} to all six nearest neighbors become equal to zero --- the so called ``magic angle'' condition.

In principle, the polycrystal/powder average also depends on the distribution of parameters $A$ and $\phi$ in Eq.(\ref{ltform}), but we found that the parameter $A$ has comparable values for all orientations and that its distribution does not add any new qualitative insight to the above discussion. Likewise, we were not able to find any particularly important aspect associated with the distribution of $\phi$, apart from the observation that it makes the frequency $\omega$ of the intermediate section fit for CaF$_2$ powder smaller than the minimal value of $\omega$ for individual single crystallites.

\section{Conclusions}
\label{conclusions}

We have presented first principles FID calculations for the powder of CaF$_2$ and for polycrystalline solid xenon. The long-time FID decay for powders/polycrystals is the superposition of the long time decays for individual single crystallites. The typical single crystallite values of the  long-time parameter $\gamma$  control the middle section of the resulting FIDs, whereas the true long-time behavior is controlled by single crystallites with the smallest value of $\gamma$. We have found that the single crystallite parameters $\gamma$ and $\omega$ are rather broadly distributed CaF$_2$, and as a result, the intermediate section beats become washed out and relatively quickly evolve to the asymptotic long-time behavior. Such a behavior might be observable in the future CaF$_2$ powder experiments with improved signal-to-noise ratio.  On the contrary,  in the case of solid xenon, the single crystallite values are more clustered, and as a result the middle section is characterized by well defined beat frequency and exponential decay constant over several orders of magnitude, while the true long-time behavior appears only at relatively later times.  We explain the above clustering of parameters $\gamma$ and $\omega$ by the more isotropic character of the fcc lattice in comparison with the simple cubic lattice. Our findings suggests that the experiments conducted so far in solid xenon  have been able to access only the intermediate section of the powder/polycrystalline FIDs, and hence observed the well-defined behavior~(\ref{ltform}). 

It is clear that, although observing well-defined behavior (\ref{ltform}) in the intermediate FID section requires suitable crystal structures,  such a behavior would be extremely unlikely, if the long-time behavior of single crystallites were different from (\ref{ltform}). Therefore, the experiments accessing the intermediate section of FIDs in polycrystalline fcc solids are appropriate to test the theoretical long-time predictions\cite{Fine-04,Fine-05} originally made mostly for single crystals. As discussed in Ref.\cite{Fine-05}, the same conclusion is likely true for solids with disordered arrangements of magnetic nuclear sites, but further experimental and theoretical investigation of this situation is necessary.


\begin{thebibliography}{19}
\bibitem[{Pre()}]{Prefactor}
\bibinfo{note}{The prefactor $2/7$ in Eq.(\ref{M2f}) follows from the
  considerations of Ref.\cite{Fine-00}. In Ref.~\cite{Fine-97}, this prefactor
  was $2/9$. The effect of this difference is very small for all calculations
  presented in this paper.}
\end{thebibliography}

\end{document}